# Organic Topological Insulators in Organometallic Lattices

Z. F. Wang, Zheng Liu and Feng Liu[*]

Department of Materials Science and Engineering, University of Utah, Salt Lake City, UT 84112, USA

[*]Correspondence and requests for materials should be addressed to F.L. (email:fliu@eng.utah.edu)

**Topological insulators (TIs) are a recently discovered class of materials having insulating bulk electronic states but conducting boundary states distinguished by nontrivial topology. So far, several generations of TIs have been theoretically predicted and experimentally confirmed, all based on inorganic materials. Here, based on first-principles calculations, we predict a family of two-dimensional organic TIs made of organometallic lattices. Designed by assembling molecular building blocks of triphenyl-metal compounds with strong spin-orbit coupling into a hexagonal lattice, this new classes of organic TIs are shown to exhibit nontrivial topological edge states that are robust against significant lattice strain. We envision that organic TIs will greatly broaden the scientific and technological impact of TIs.**

The concept of topological order in condensed matter physics provides a new perspective for understanding the origin of different quantum phases and has generated intense recent interest in searching for nontrivial topological materials, so called topological insulators (TIs)[1-21]. The defining signature of a TI is the existence of robust conducting edge or surface states on the boundary of normal insulators. These characteristic boundary states have a topological origin, which are protected from elastic backscattering and localization, and hence hold potential for applications in spintronics and quantum computation devices.

So far, several generations of TI have been predicted theoretically and confirmed experimentally. The first one is two-dimensional (2D) TI (quantum spin Hall system) in an HgTe based quantum well[5,6]. Subsequently, several binary ($Bi_2Te_3$, $Bi_2Se_3$, and $Sb_2Te_3$)[7-10], ternary ($TlBiTe_2$, $TlBiSe_2$ and $PbBi_2Te_4$)[11-13] and Heusler compounds[14-16] have been found to be three-dimensional (3D) TIs. Most recently, the Bi bilayer[17-20] and actinide based binary compounds[21] have been predicted to be new families of 2D and 3D TIs, respectively. However, these existing families of TIs are all inorganic materials. On the other hand, many conventional inorganic materials and devices have later found their organic counterparts, such as organic superconductors[22], light-emitting diodes[23], solar cells[24] and field-effect transistors[25]. In general, organic counterparts of inorganic materials have the added advantages of low cost, easy fabrication and mechanical flexibility. Therefore, an interesting question is whether an organic TI (OTI) exists?

TIs are distinguished from ordinary insulators by nontrivial topological invariants associated with bulk electronic structure. In particular, their electronic structures are characterized by a bulk spin-orbit coupling (SOC) gap and an odd number of Dirac-like edge states for a 2D TI (or surface states for a 3D TI) connecting the conduction and valence edge at certain k-points; the latter can be guaranteed by band inversion in an inversion symmetric system that gives rise to an odd total parity of all time reversal invariant points[2]. To design or search for a TI material, two key ingredients are essential: one is the lattice symmetry and the other is the SOC. The first condition is readily met by a single-band model on a hexagonal lattice, such as graphene[1], which is predicted to be a 2D TI when a SOC gap is present. However, the SOC in graphene is too small to open a gap at finite temperature[26], and proposals have been made to extrinsically enhance SOC of graphene by doping with heavy atoms[27,28].

The emerging theme of materials genome inspires exciting opportunities to computationally design novel materials with desired properties. Meanwhile, advances on synthetic chemistry and nanotechnology have shown the potential in producing complex 2D lattices, *i.e.* covalent organic frameworks. Keeping the above two ingredients in mind, we have computationally designed a new family of OTIs. We have looked into organometallic compounds as molecular building blocks, which consist of C-C and C-metal bonds. To satisfy the first condition, we narrowed down our choices to the triphenyl-metal compounds with one metal atom bonded symmetrically to three benzene rings, so that they can be readily self-assembled into a hexagonal lattice, as shown in Fig. 1. To satisfy the second condition, we simply choose heavy metal atoms, such as Pb and Bi, with strong SOC.

**Results**

**2D triphenyl-lead lattice.** We first consider the triphenyl-lead (TL) [Pb($C_6H_5$)$_3$] lattice. The TL molecule consists of a Pb atom bonded with three benzene rings with three-fold rotational symmetry (Fig. 1a). When bridged with a Pb atom (Fig. 1b), they naturally form a 2D hexagonal lattice (Fig. 1c). There are two Pb atoms (labeled as 1 and 2 in Fig. 1d) and three benzene rings with a chemical formula of $Pb_2C_{18}H_{12}$ in each unit cell and the neighboring benzene rings are bridge-bonded through the para-Pb atoms. The 2D lattice is slightly buckled with the para-Pb atoms moving alternately up and down out of the plane of benzene rings. This is because the Pb has a $6s^26p^2$ electronic configuration that favors the $sp^3$ hybridization. The equilibrium lattice constant is found to be 12.36 Å, with the Pb-Pb distance (*l*) and height difference (*h*) being 7.46 Å and 2.18 Å, respectively, as indicated in Fig. 1b. The comparison of lattice energy between the flat and buckled structure and the stability analysis of the TL lattice are shown in Supplementary Fig. S1 and S2.

Without SOC, the electronic band structure of TL lattice displays a Dirac cone at the K point, with the Fermi level located exactly at the Dirac point, as shown in Fig. 2a and 2b. Turning on the SOC, a gap of ~8.6 meV opens at the K point, as shown in Fig. 2c and 2d. We label this SOC induced gap as $E_g^D$, to be distinguished from the conventional insulator band gap. For a 2D TI, there should be an odd number of Dirac-like edge states connecting the conduction and valence edge at certain k-points. Thus, we have checked the existence of nontrivial topological edge states of the TL lattice, using the Wannier90 package[29]. We fit a tight-binding Hamiltonian with maximally localized Wannier functions (MLWFs) to the first-principles band structures. With these MLWFs, the edge Green's function[30] of the semi-infinite TL lattice is constructed and the local density of state (LDOS) of edges is calculated. This method provides a direct connectivity between the edge states and the bulk states. In Fig. 2e, a comparison between the Wannier and first-principles bands is given around the Dirac-cone gap, showing very good agreement. The LDOS of the zigzag edge is shown in Fig. 2f, where one can clearly see the nontrivial topological gapless edge states that connect the band edge states between K and K′ and form a one-dimensional (1D) Dirac cone at the boundary of the Brillouin zone (BZ), signifying our designed TL lattice to be a 2D TI.

We note that because the Dirac bands have a relative small band width (~ 1eV), there could be other instabilities of gap opening mechanisms, such as those associated with magnetism, dimerization and charge density wave. These can be interesting topics of future studies but beyond the scope of current work within the framework of density functional theory (DFT). Based on our DFT calculations, the system remains always a non-magnetic ground state.

**Effective Hamiltonian.** Because the topological nature of the TL lattice can be determined by the physics near the K (K′) point, it is possible to write down an effective Hamiltonian to characterize its low-energy properties (see Supplementary Fig. S3 and S4, and Supplementary Note 1). Similar to the silicene model[31,32,33], which is also predicted to be a 2D TI, the effective Hamiltonian for the TL lattice around the K and K′ can be written as

$$H_\eta = \hbar v_F (k_x \tau_x - \eta k_y \tau_y) + \eta \tau_z h_{11} + \varepsilon_0 I \qquad (1a),$$

$$h_{11} = -\lambda_{SO}\sigma_z - a\lambda_R(k_y\sigma_x - k_x\sigma_y) \qquad (1b),$$

where $\eta= \pm 1$ corresponds to K and K′ points, respectively, $\sigma$ ($\tau$) is the Pauli matrix of spin (sublattice), $v_F$ is the Fermi velocity, $\varepsilon_0$ is the on-site energy, $I$ is the 4×4 identity matrix, $\lambda_{SO}$ is the effective SOC, $\lambda_R$ is the Rashba SOC and $a$ is lattice constant. A comparison between the first-principles and effective-Hamiltonian band structure is shown in Fig. 2b and 2d. The corresponding fitting parameters are $\varepsilon_0$=0 eV, $\hbar v_F$=1.13 eV·Å, $\lambda_{SO}$=0.0043 eV, $a\lambda_R$=0.316 eV·Å. We note that there is a non-zero Rashba SOC in our model due to structural buckling, which is absent in the flat graphene structure. Buckling makes the two sublattices of metal atoms in the different plane to have both in-plane and out-of-plane SOC components[32]. The in-plane component, like in graphene, is responsible for gap opening. The out-of-plane component, resulting from the potential gradient in $z$-direction[32] (absent in graphene), changes slightly the band dispersion (slope) around the Dirac point, but not important to topology. So the Rashba SOC here is intrinsic different from the extrinsic Rashba SOC induced by breaking of the inversion symmetry.

**2D triphenyl-bismuth lattice.** The TL lattice represents a "true" TI having the right band filling with the Fermi level located exactly in the middle of the Dirac-cone gap (or Dirac point without SOC), but the gap is relatively small, $E_g^D$ ~ 8.6 meV. One may choose a different metal atom with larger SOC to increase this gap. For example, we have calculated the electronic and topological properties of triphenyl-bismuth (TB) [Bi(C$_6$H$_5$)$_3$] lattice by replacing Pb with Bi, as shown in Fig. 3. The stability and structural properties of TB lattice are similar to the TL lattice, as shown in Supplementary Fig. S1. It has all the required electronic and topological properties for a TI and a Dirac-cone gap as large as ~43 meV.

However, its Fermi level is ~0.31 eV above the Dirac point without SOC (Fig. 3a) because Bi has two extra electrons than Pb. Thus, to truly make the TB lattice a TI material, one must move the Fermi level into the Dirac-cone gap by doping. This requires doping of two holes per unit cell, because there are two degenerate bands in between the Fermi level and Dirac-cone gap (see Supplementary Fig. S5 and Supplementary Note 2). This can possibly be achieved by gating effect. It is worth noting that the OTIs we demonstrate here represent a real material system that realizes the Kane-Mele model[1] of 2D TI with a sizable gap, comparable to the known inorganic 2D TI of HgTe quantum well[5,6].

**Strain effects.** For an inversion symmetric system, its topology can also be identified by band inversion that gives rise to an odd total parity of all time reversal invariant points[2], via calculation of topological index $Z_2$ number. For the lattices we consider here, there are four time-reversal invariant k-points, one Γ and three M's in the BZ. We have used this method to also check the strain tolerance of the nontrivial topology of the TL and TB lattices, without repeating the calculation of edge state as a function of strain. Since the proposed lattices are likely to be assembled on the substrate, they may not maintain their free-standing lattice constant. We found that the SOC gap ($E_g^D$) varies only slightly with the in-plane biaxial strain, and there is no gap closing even for strains up to ±10%. This can be reasonably understood because the strength of SOC is a local property associated with metal atom (Pb vs. Bi) rather insensitive to strain. The calculated $Z_2$ number is one within the range of ±10% strain applied for both TL and doped TB lattice. This means that the TL and TB lattices maintain a nontrivial topology (for a trivial topological phase, $Z_2$ =0), which is very stable against strain.

Since the topological invariance is a global property in the whole BZ of the lattice, it is expected to be robust under uniform lattice deformation, making it easier for experimental realization and characterization.

**Discussion**

Lastly, we comment on the experimental feasibility to produce and characterize our proposed 2D OTIs. Organometallic chemistry and substrate-mediated molecular self-assembly are both well-established methods for producing exotic materials. Most importantly, recent developments in this field have already successfully synthesized the oriented 2D covalent organic frameworks (COFs) involving directional metal-C, C-C as well as other organic bonds[34-39], which are exactly what needed for producing our structure. These existing COFs are produced with well-defined lattice symmetry (including the needed hexagonal symmetry) and long-range periodicity, as well as good uniformity in large sample size. (see, e.g., scanning tunneling microscope and low energy electron diffraction images in Fig. 2 of Ref. 39). It is also worth noting that these COFs can be experimentally made with different metal atoms (such as Fe, Cu and Co in Refs. 38 and 39) and variety of molecules, while all the metals and molecular precursors needed in our proposed structures are commercially available and have been used in organometallic chemistry and molecular self-assembly before. Therefore, we believe the existing experimental results of COFs have clearly demonstrated the feasibility of producing our proposed OTIs with desired lattice symmetry, good crystalline quality and sufficient size.

**Methods**

DFT calculations for the stability, band structure and topological properties of 2D organometallic lattices made of triphenyl-lead and triphenyl-bismuth were carried out in the framework of the Perdew-Burke-Ernzerhof type generalized gradient approximation functional using the VASP package[40]. All self-consistent calculations were performed with a plane-wave cutoff of 600 eV on a 7×7×1 Monkhorst-Pack k-point mesh on supercells with a vacuum layer more than 15 Å thick to ensure decoupling between neighboring slabs. For structural relaxation, all the atoms are allowed to relax until atomic forces are smaller than 0.01 eV/Å.

**Reference**

1. Kane, C. L. & Mele, E. J. Quantum spin hall effect in graphene. *Phys. Rev. Lett.* **95**, 226801 (2005).

2. Fu, L. & Kane, C. L. Topological insulators with inversion symmetry. *Phys. Rev. B* **76**, 045302 (2007).

3. Hasan, M. Z. & Kane, C. L. Colloquium: Topological insulators. *Rev. Mod. Phys.* **82**, 3045-3067 (2010).

4. Qi, X.-L. & Zhang, S.-C. Topological insulators and superconductors. *Rev. Mod. Phys.* **83**, 1057-1110 (2011).

5. Bernevig, B. A. *et al.* Quantum spin hall effect and topological phase transition in HgTe quantum wells. *Science*. **314**, 1757-1761 (2006).

6. König, M. *et al.* Quantum spin hall insulator state in HgTe quantum wells. *Science* **318**, 766-770 (2007).

7. Hsieh, D. *et al.* A topological Dirac insulator in a quantum spin hall phase. *Nat.* **452**, 970-974 (2008).

8. Xia, Y. *et al.* Observation of a large-gap topological-insulator class with a single Dirac cone on the surface. *Nat. Phys.* **5**, 398-402 (2009).

9. Zhang, H. J. *et al.* Topological insulators in $Bi_2Se_3$, $Bi_2Te_3$ and $Sb_2Te_3$ with a single Dirac cone on the surface. *Nat. Phys.* **5**, 438-442 (2009).

10. Chen, Y. L. *et al.* Experimental realization of a three-dimensional topological insulator, $Bi_2Te_3$. *Science* **325**, 178-181 (2009).

11. Lin, H. *et al.* Single-dirac-cone topological surface states in the $TlBiSe_2$ class of

topological semiconductors. *Phys. Rev. Lett.* **105**, 036404 (2010).

12. Chen, Y. L. *et al.* Single dirac cone topological surface state and unusual thermoelectric property of compounds from a new topological insulator family. *Phys. Rev. Lett.* **105**, 266401 (2010).

13. Souma, S. *et al.* Topological surface states in lead-based ternary telluride $Pb(Bi_{1-x}Sb_x)_2Te_4$. *Phys. Rev. Lett.* **108**, 116801 (2012).

14. Xiao, D. *et al.* Half-heusler compounds as a new class of three-dimensional topological insulators. *Phys. Rev. Lett.* **105**, 096404 (2010).

15. Lin, H. *et al.* Half-heusler ternary compounds as new multifunctional experimental platforms for topological quantum phenomena. *Nat. Mater.* **9**, 546-549 (2010).

16. Chadov, S. *et al.* Tunable multifunctional topological insulators in ternary heusler compounds. *Nat. Mater.* **9**, 541-545 (2010).

17. Murakami, S. Quantum spin hall effect and enhanced magnetic response by spin-orbit coupling. *Phys. Rev. Lett.* **97**, 236805 (2006).

18. Hirahara, T. *et al.* Interfacing 2D and 3D topological insulators: Bi(111) bilayer on $Bi_2Te_3$. *Phys. Rev. Lett.* **107**, 166801 (2011).

19. Liu, Z. *et al.* Stable nontrivial $Z_2$ topology in ultrathin Bi (111) Films: a first-principles study. *Phys. Rev. Lett.* **107**, 136805 (2011).

20. Yang, F. *et al.* Spatial and energy distribution of topological edge states in single Bi(111) bilayer. *Phys. Rev. Lett.* **109**, 016801 (2012).

21. Zhang, X. *et al.* Actinide topological insulator materials with strong interaction. *Science* **335**, 1464-1466 (2012).

22. Jérome, D. *et al.* Superconductivity in a synthetic organic conductor (TMTSF)$_2$PF$_6$. *J. Phys. Lett. (Paris)* **41**, L95-98 (1980).

23. Tang, C. W. & Vanslyke, S. A. Organic electroluminescent diodes. *Appl. Phys. Lett.* **51**, 913 (1987).

24. Kearns, D. & Calvin, M. Photovoltaic effect and photoconductivity in laminated organic systems. *J. Chem. Phys.* **29**, 950 (1958).

25. Koezuka, H., Tsumura, A. & Ando, T. Field-effect transistor with polythiophene thin film. *Synth. Met.* **18**, 699-704 (1987).

26. Min, H. *et al.* Intrinsic and Rashba spin-orbit interactions in graphene sheets. *Phys. Rev. B* **74**, 165310 (2006).

27. Weeks, C. *et al.* Engineering a robust quantum spin hall state in graphene via adatom deposition. *Phys. Rev. X* **1**, 021001 (2011).

28. Qiao, Z. *et al.* Two-dimensional topological insulator state and topological phase transition in bilayer graphene. *Phys. Rev. Lett.* **107**, 256801 (2011).

29. Mostofi, A. A. *et al.* Wannier90: a tool for obtaining maximally-localized wannier functions. *Comput. Phys. Commun.* **178**, 685-699 (2008).

30. Sancho, M. P. L. *et al.* Highly convergent schemes for the calculation of bulk and surface Green functions. *J. Phys. F* **15**, 851 (1985).

31. Liu, C.-C. *et al.* Quantum spin hall effect in silicene and two-dimensional germanium. *Phys. Rev. Lett.* **107**, 076802 (2011).

32. Liu, C.-C. *et al.* Low-energy effective Hamiltonian involving spin-orbit coupling in silicene and two-dimensional germanium and tin. *Phys. Rev. B* **84**, 195430 (2011).

33. Ezawa, M. A topological insulator and helical zero mode in silicene under an inhomogeneous electric field. *New. J. Phys.* **14**, 033003 (2012).

34. Sakamoto, J. *et al.* Two-dimensional polymers: just a dream of synthetic chemists? *Angew. Chem. Int. Ed.* **48**, 1030-1069 (2009).

35. Grill, L. *et al.* Nano-architectures by covalent assembly of molecular building blocks. *Nat. Nanotech.* **2**, 687-691 (2007).

36. Côté, A. P. *et al.* Porous, crystalline, covalent organic frameworks. *Science* **310**, 1166-1170 (2005).

37. Colson, J. W. *et al.* Oriented 2D covalent organic framework thin films on single-layer graphene. *Science* **332**, 228-231 (2011).

38. Shi, Z. *et al.* Thermodynamics and selectivity of two-dimensional metallo-supramolecular self-assembly resolved at molecular scale. *J. Am. Chem. Soc.* **133**, 6150–6153 (2011).

39. Schlickum, U. *et al.* Metal-organic honeycomb nanomeshes with tunable cavity size. *Nano Lett.* **7**, 3813-3817 (2007).

40. Kresse, G. & Hafner, J. Ab initio molecular dynamics for liquid metals. *Phys. Rev. B* **47**, 558-561 (1993).

**Acknowledgements**

We thank C.X. Liu for helpful discussion. This work was supported by US DOE-BES (Grant No. DE-FG02-04ER46148). Z.F.W. also acknowledges support from ARL (Cooperative Agreement No. W911NF-12-2-0023). We thank the CHPC at University of Utah and NERSC for providing the computing resources.

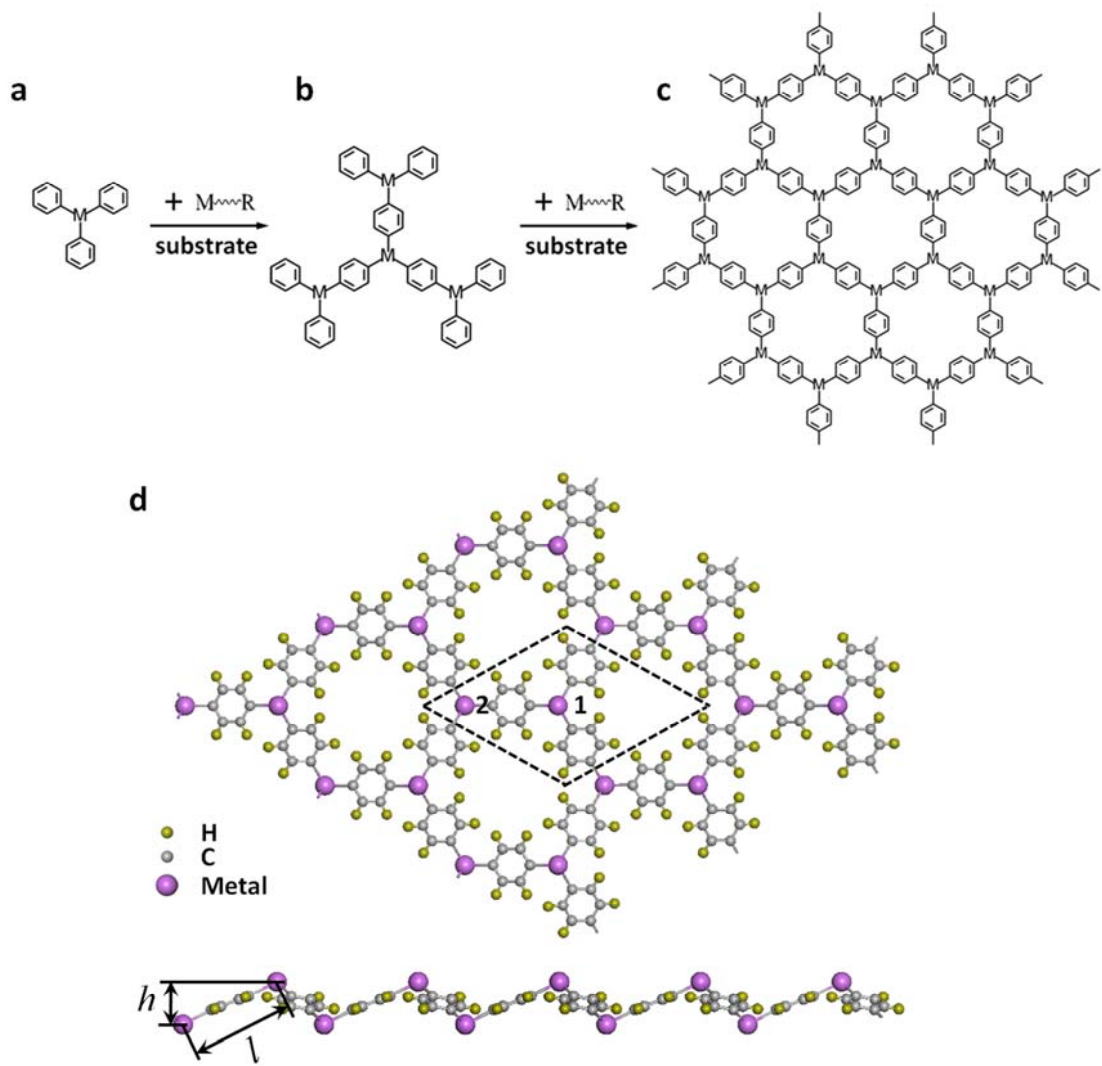

**Figure 1. Schematics of 2D organometallic lattice.** (a)-(c) The proposed synthesis process from the triphenyl-metal molecules to the 2D organometallic lattices. M is the metal and R is the functional group attached to the metal atom. (d) Top and side view of the 2D organometallic superlattice. Dashed lines show the unit cell and the two metal atoms are labeled with 1 and 2. *l* and *h* are the distance and height difference between the two metal atoms.

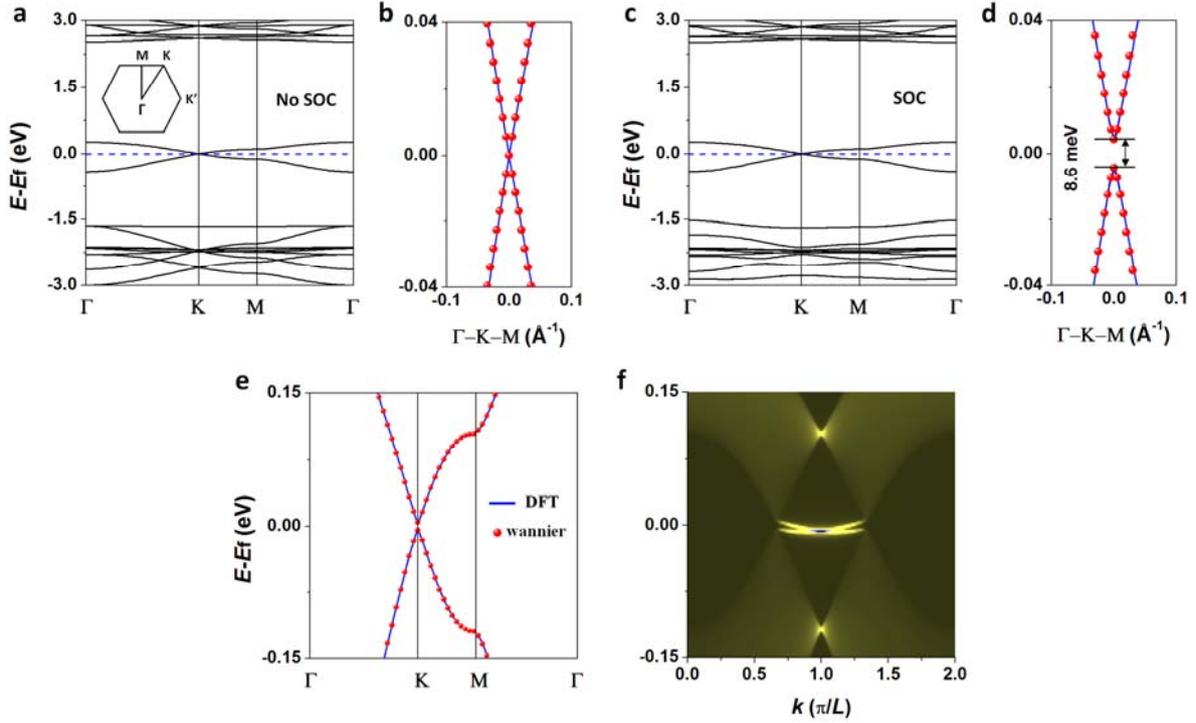

**Figure 2. Topological properties of 2D TL lattice.** (a) and (c) Band structures of TL lattice without and with SOC, respectively. The dashed line indicates the Fermi level. The inset is the 2D first Brillouin zone, and Γ, K, M and K′ are the high symmetric k-points. (b) and (d) The zoom-in band structures around the Dirac point without and with SOC, respectively. The red dots are the fitted bands using the effective Hamiltonian. (e) Wannier and first-principles band structures around the Dirac-cone gap. (f) Energy and momentum dependent LDOS of the zigzag edge of the semi-infinite TL lattice, $L$ is the unit cell length along the edge.

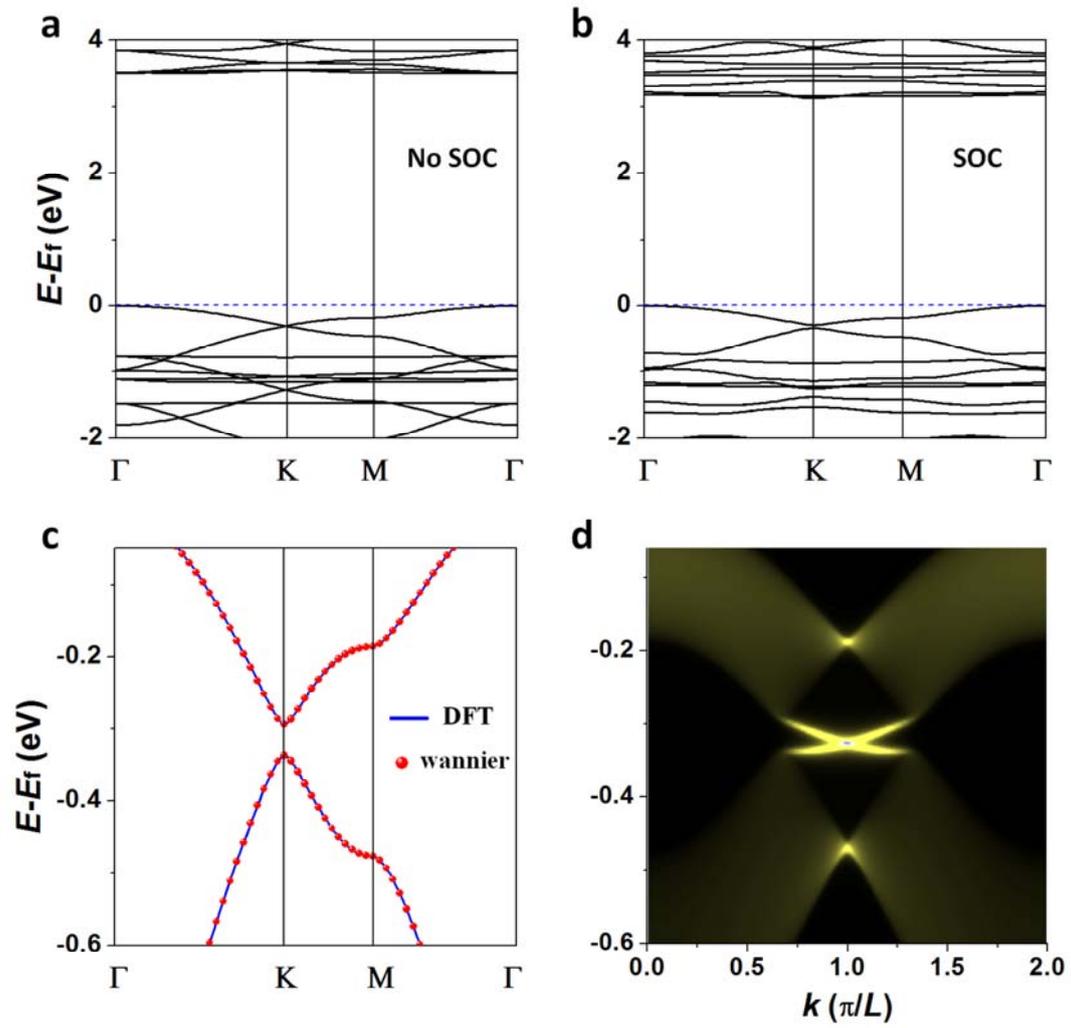

**Figure 3. Topological properties of 2D TB lattice.** (a) and (b) Band structures of TB lattice without and with SOC, respectively. The dashed line indicates the Fermi level. (c) Wannier and first-principles band structures around the Dirac-cone gap. (d) Energy and momentum dependent LDOS of the zigzag edge of the semi-infinite TB lattice, $L$ is the unit cell length along the edge.